\documentclass[aps,prl,twocolumn,showpacs,floatfix,nobibnotes,superscriptaddress,longbibliography,nofootinbib]{revtex4-1}

\usepackage[utf8]{inputenc}
\usepackage{amssymb}
\usepackage{graphicx}
\usepackage{verbatim}
\usepackage{epsfig}
\usepackage{bm}
\usepackage{color}
\usepackage{xcolor}
\usepackage{float}
\usepackage{dcolumn}
\usepackage{multirow} 
\usepackage{physics}
\usepackage{braket}
\usepackage{ytableau}
\usepackage{footnote}
\usepackage{amsmath}
\usepackage[unicode=true,
  linktocpage,
  linkbordercolor={0.5 0.5 1},
  citebordercolor={0.5 1 0.5},
  colorlinks=true,
  linkcolor=blue,
  citecolor=blue,
  urlcolor=blue]{hyperref}
\usepackage{soul}
\setstcolor{purple}
\soulregister{\cite}{7}

\begin{document}
\title{Emergent Wigner \texorpdfstring{SU(4)}{SU(4)} Symmetry and Electroweak Responses of Light Nuclei}

\author{Simone Salvatore Li Muli}
\email{slimuli@utk.edu}
\affiliation{Department of Physics and Astronomy, Chalmers University of Technology, SE-412 96 G\"oteborg, Sweden}
\affiliation{Department of Physics and Astronomy, University of Tennessee, Knoxville, TN 37996, USA}

\author{Christian Forssén}
\email{christian.forssen@chalmers.se}
\affiliation{Department of Physics and Astronomy, Chalmers University of Technology, SE-412 96 G\"oteborg, Sweden}

\author{Daniel R. Phillips}
\email{phillid1@ohio.edu}
\affiliation{Department of Physics and Astronomy and Institute of Nuclear and Particle Physics, Ohio University, Athens, Ohio 45701, USA}
\affiliation{Department of Physics and Astronomy, Chalmers University of Technology, SE-412 96 G\"oteborg, Sweden}

\begin{abstract}
Emergent Wigner's $\mathrm{SU}(4)$ symmetry relates weak axial-vector and electromagnetic nuclear responses, since the isovector charge, Gamow--Teller, and spin-$M1$ one-body operators form one $\mathrm{SU}(4)$ tensor. 
For exact symmetry, these form factors share a single shape and their $q=0$ ratios are determined by a generalized Wigner--Eckart theorem.  
Realistic \textit{ab initio} calculations for $A = 3$, $6$, and $7$ nuclei confirm this up to momentum transfer $q=400$~MeV/$c$, with symmetry breaking that we quantify. In particular, at the one-body level the $^6$He Gamow--Teller radius exceeds the $^6$Li magnetic radius by only $(4\pm2)\%$. Electromagnetic data can therefore calibrate axial responses at finite $q$, where direct data are scarce. For instance, electron scattering on $^6$Li can constrain the $^6$He $\beta$ spectrum used in searches for physics beyond the Standard Model.
\end{abstract}

\maketitle

\textit{Introduction}---The conserved-vector-current (CVC) hypothesis~\cite{fey58} placed the weak vector and the electromagnetic isovector currents in the same isospin triplet, allowing vector weak transitions to be inferred from electromagnetic data. However, CVC provides no analogous constraint for the axial current. At the one-body level, isospin relates Gamow--Teller and isovector spin-$M1$ (the spin component of the magnetic dipole operator) transitions, but only when they involve isobaric analog states.

In this Letter, we show that a larger symmetry extends this relation from isospin multiplets to $\mathrm{SU}(4)$ multiplets. Wigner's $\mathrm{SU}(4)$ spin-isospin symmetry~\cite{wig37} has been demonstrated to emerge in \textit{ab initio} predictions for light nuclei~\cite{lim26_1,dan26,lin23,ngu24,lin25,lin26}. 
Reference~\cite{lim26_1} exploited $\mathrm{SU}(4)$ selection rules for Gamow--Teller strengths at $q=0$ to explain the $\beta$-decay patterns for $A=3$-$8$ systems. 
At $q>0$, however, the physical currents are no longer $\mathrm{SU}(4)$ generators, so these selection rules need not hold.

\begin{figure}[h]
\centering
\includegraphics[width=\columnwidth]{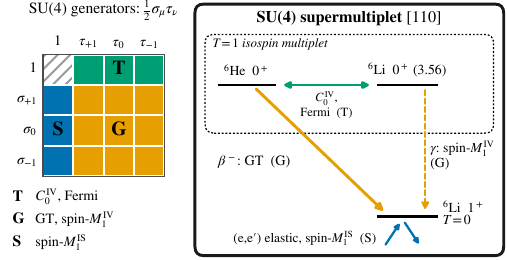}
\caption{Left: the 15 one-body $\mathrm{SU}(4)$ generators $\tfrac{1}{2}\sigma_\mu$, $\tfrac{1}{2}\tau_\nu$, and $\tfrac{1}{2}\sigma_\mu\tau_\nu$ (spherical components; row $\times$ column), grouped into the sectors $T$, $S$ and $G$ with their observables; the hatched cell is the scalar. Right: the $[110]$ supermultiplet of $A=6$. The dotted box denotes the $T=1$ analog states. Isospin relates  transitions within this multiplet (green arrows, $T$).
Isospin also imposes separate relations on transitions between it and the ${}^6$Li ground state (orange arrows, $G$). But $\mathrm{SU}(4)$ connects these two otherwise independent transition sets to one another and also to the isoscalar spin response of $^6$Li (blue arrows, $S$).}
\label{fig:schematic}
\end{figure}

Nevertheless, components of the one-body Gamow--Teller (GT), isoscalar (IS) and isovector (IV) spin-magnetic-dipole ($M_1^{\mathrm{IS}}$, $M_1^{\mathrm{IV}}$), and isovector charge-monopole ($C_0^{\mathrm{IV}}$) operators form a single $\mathrm{SU}(4)$ adjoint tensor at any momentum transfer. Within a supermultiplet, their form factors are then governed by one symmetry-determined function of $q$, which we dub the $\mathrm{SU}(4)$ adjoint form factor. Their normalizations at $q=0$, where the operators reduce to SU(4) generators, are fixed by group theory alone. In this Letter we show that \textit{ab initio} calculations of light nuclei produce form factors that obey these $\mathrm{SU}(4)$ symmetry relations up to corrections that are typically small. The emergent Wigner symmetry links one-body axial and electromagnetic form factors, and thus radii, \emph{across} isospin multiplets, beyond the reach of both CVC and isospin.
Figure~\ref{fig:schematic} illustrates this in the $A=6$ supermultiplet: isospin relates the $^6$He Gamow--Teller transition only to the spin-$M_1^{\mathrm{IV}}$ decay of its isobaric analog state in $^6$Li, whereas $\mathrm{SU}(4)$ also ties it to the $C_0^{\mathrm{IV}}$ form factor of $^6$He, the Fermi $^6$He$\to{}^6$Li$^*$(3.56~MeV) transition, and to the spin-$M_1^{\mathrm{IS}}$ response of the $^6$Li ground state. Analogous relations hold in every supermultiplet, e.g., the $[111]$ mirror pairs of $A=3$ and $7$.

\textit{Theoretical Framework}---Wigner's $\mathrm{SU}(4)$ group acts on the four-dimensional spin-isospin space of the nucleon~\cite{wig37,wig39}. Its fifteen generators, $S_\mu = \tfrac{1}{2}\sum_k \sigma_\mu(k)$, $T_\nu = \tfrac{1}{2}\sum_k \tau_\nu(k)$, and $G_{\mu\nu} = \tfrac{1}{2}\sum_k \sigma_\mu(k)\tau_\nu(k)$, written in spherical components $\mu,\nu\in\{0,\pm1\}$ [$\sigma_{\pm1}=\mp(\sigma_x\pm i\sigma_y)/\sqrt{2}$, $\sigma_0=\sigma_z$, and likewise for $\tau$], transform as the adjoint representation $[211]$, which decomposes under the subgroup chain $\mathrm{SU}(4) \supset \mathrm{SU}(2)_S \otimes \mathrm{SU}(2)_T$ into three sectors, labeled by their generators: $[211] \;\longrightarrow\; \mathrm{S}\,(S_{\rm op}{=}1,T_{\rm op}{=}0) \;\oplus\; \mathrm{T}\,(0,1) \;\oplus\; \mathrm{G}\,(1,1)$. 
Generators cannot connect distinct irreducible representations (irreps), so the corresponding matrix elements vanish between different supermultiplets~\cite{wig39,nan72,muk74,vog93}. Within a supermultiplet $[f]$, the algebra fixes the matrix elements of all three sectors relative to one another.
The 15 generators can be mapped to long-wavelength limits of several electroweak one-body operators. The $\mathrm{S}$ generators map to the spin-$M_1^{\mathrm{IS}}$, the neutral ($\nu=0$) and charge-changing ($\nu=\pm1$) components of the $\mathrm{T}$ generators are related to the $C_0^{\mathrm{IV}}$ (and thus to the Fermi operators), while the neutral and charge-changing components of the $\mathrm{G}$ generators are related to spin-$M_1^{\mathrm{IV}}$ and GT operators.

To extend this link to finite momentum transfer $q$, we define the $\mathrm{SU}(4)$ adjoint monopole density
\begin{align}
  \rho^{[211]}_{(S_{\rm op}\mu)(T_{\rm op}\nu)}(q) &\equiv \sum_k j_0(qr_k)\,Y_{00}(\hat{\bm{r}}_k)\,\Gamma^{[211]}_{(S_{\rm op}\mu)(T_{\rm op}\nu)}(k), \label{eq:adjoint}
\end{align}
where $\bm{r}_k$ is the coordinate of nucleon $k$ and $\Gamma^{[211]}$ is the one-nucleon $\mathrm{SU}(4)$ adjoint tensor, labeled by its spin and isospin ranks and projections in the standard phases of Ref.~\cite{hec69} (End Matter): $\Gamma^{[211]}_{(1\mu)(00)}=\tfrac{1}{2}\sigma_\mu$, $\Gamma^{[211]}_{(00)(1\nu)}=\tfrac{1}{2}\tau_\nu$, and $\Gamma^{[211]}_{(1\mu)(1\nu)}=-\tfrac{1}{2}\sigma_\mu\tau_\nu$. The components of Eq.~\eqref{eq:adjoint} are the monopole Fourier transforms of the spin, isospin, and spin-isospin densities: their $q\to0$ limits are proportional to the generators $S_\mu$, $T_\nu$, and $G_{\mu\nu}$, and their $q$ dependence describes how these $\mathrm{SU}(4)$ charges are distributed radially in the nucleus. Although they are no longer generators for $q>0$, they remain components of a single $\mathrm{SO}(3)_L\otimes \mathrm{SU}(4)$ tensor of orbital rank $\mathcal{L}=0$ and $\mathrm{SU}(4)$ type $[211]$. 

We define form factors $F^{[211]}_X(q)$ as $\mathrm{SO}(3)$ reduced matrix elements of the components $\rho^{[211]}_{(S_{\rm op}\mu)(T_{\rm op}\nu)}(q)$ in the spin, isospin, and spin-isospin sectors of $\mathrm{SU}(4)$:  $X\in\{\mathrm{S},\mathrm{T},\mathrm{G}\}$. The form factors are defined for transitions between nuclear states of good $L$, $S$, and $T$ that lie in particular $\mathrm{SU}(4)$ supermultiplets $[f_i]$ and $[f_f]$. 
The Generalized Wigner--Eckart (GWE) theorem for $\mathrm{SO}(3)_L\otimes \mathrm{SU}(4)$ then implies that they obey:
\begin{align}
\big \langle [f_f]&S_fT_fM_{T_f}\,L_f\,J_f\|\rho^{[211]}_{(S_{\rm op})(T_{\rm op}\nu)}(q)\|[f_i]S_iT_iM_{T_i}\,L_i\,J_i\big\rangle
\nonumber \\
  &\quad = [J_f] \times \mathrm{CG}_T \times \mathcal{G} \times W \times \mathcal{R}(q),
  \label{eq:gwe}
\end{align}
with $[J]\equiv\sqrt{2J+1}$. Here we have used the convention of Edmonds~\cite{edm57} for the $\mathrm{SO}(3)$ algebra, and 
\begin{equation}
  \mathcal{G} = [S_f][L_f][J_i][S_{\rm op}]
  \begin{Bmatrix} S_i & L_i & J_i \\ S_{\rm op} & 0 & S_{\rm op} \\ S_f & L_f & J_f \end{Bmatrix}
  \label{eq:geom}
\end{equation}
is a geometric factor.
$\mathrm{CG}_T=\langle T_iM_{T_i};T_{\rm op}\nu|T_fM_{T_f}\rangle$ is the isospin Clebsch--Gordan coefficient, and $W \equiv \langle [f_i]S_iT_i;[211]S_{\rm op}T_{\rm op}\|[f_f]S_fT_f\rangle$ is the reduced $\mathrm{SU}(4)$ Wigner coefficient~\cite{hec69}\footnote{A multiplicity label $\rho$ is needed if $[f_f]$ occurs more than once in $[f_i]\otimes[211]$, but this is not an issue for the cases considered here.}, with $S_{\rm op}$ and $T_{\rm op}$ the spin-isospin ranks of the operator. 

The quantity $\mathcal{R}(q)$ is then the reduced many-body matrix element with respect to $\mathrm{SO}(3)_L \otimes \mathrm{SU}(4)$ and Eq.~\eqref{eq:gwe} states that all form factors within a particular supermultiplet are proportional to it: $F^{[211]}_X(q)=c_X\,\mathcal{R}(q)$ with $c_X \equiv [J_f]\,\mathrm{CG}_T\,\mathcal{G}\,W$. The three normalized form factors $\widetilde{F}_X(q) \equiv F^{[211]}_X(q)/F^{[211]}_X(0)$ are therefore equal. We denote this single adjoint form factor by $F^{[211]}(q)$:
\begin{equation}
F^{[211]}(q) \equiv \widetilde{F}_G(q)=\widetilde{F}_S(q)=\widetilde{F}_T(q),  
  \label{eq:shapelock}
\end{equation}
and the corresponding mean-square radii are all equal. Furthermore, because the operators reduce to $\mathrm{SU}(4)$ generators at $q=0$, the normalizations $F^{[211]}_X(0)$ can be computed using only group theory (End Matter, Table~\ref{tab:su4_limits}).

Three effects separate realistic electroweak responses from the ideal limit of Eq.~\eqref{eq:shapelock}. First, Eq.~\eqref{eq:shapelock} assumes that the nuclear states connected by the different channels belong to the same supermultiplet $[f]$ and share a single orbital wave function, as they do for an $\mathrm{SU}(4)$-invariant Hamiltonian. However, realistic nuclear forces break $\mathrm{SU}(4)$. They mix higher irreps into the nuclear eigenstates and, even within the dominant irrep, produce slightly different orbital wave functions for the nuclei within a supermultiplet.

Second, at finite $q$ the physical electroweak currents are not pure $\rho^{[211]}$ tensors. Through order $|\bm{p}|/M_N$, their one-body content is carried by seven multipole operators~\cite{don79,hax08} (defined explicitly in the End Matter), which we classify by their long-wavelength behavior. \textit{Class A} operators reduce at $q=0$ to an operator proportional to a component of the adjoint tensor $\Gamma^{[211]}_{(S_{\rm op}\mu)(T_{\rm op}\nu)}$ summed over nucleons, i.e., to the many-body $\mathrm{SU}(4)$ generators. In contrast, \textit{Class B} operators have no $\mathrm{SU}(4)$ generator limit. The isoscalar parts of the charge and convection multipoles, $C^{\mathrm{IS}}_J$, $\Delta^{\mathrm{IS}}_J$, and $\Delta'^{\mathrm{IS}}_J$, carry no spin-isospin structure, and are thus $\mathrm{SU}(4)$ scalars and block-diagonal at any $q$. All others, namely the isovector convection currents $\Delta^{\mathrm{IV}}_J,\Delta'^{\mathrm{IV}}_J$, the spin-velocity operator $\Omega_J$, the transverse spin multipole $\Sigma_J$, and higher multipoles $C^{\mathrm{IV}}_{J\ge1}$, $\Sigma'_{J\ge2}$, $\Sigma''_{J\ge2}$, become a product of a non-trivial orbital operator and a component of $\Gamma^{[211]}$ as $q \rightarrow 0$. They therefore connect different supermultiplets.

Third, even the \textit{Class A} spin $J=1$ operators are not pure components of the adjoint monopole density. $\Sigma'_1$ and $\Sigma''_1$ are orthogonal combinations of an $\mathcal{L}=0$ term, which is a component of $\rho^{[211]}$, and an $\mathcal{L}=2$ term (End Matter), each with its own reduced matrix element. Only the $\mathcal{L}=0$ follows $F^{[211]}$.
However, we show below that in light nuclei the $\mathcal{L}=2$ term is  suppressed: it vanishes between $L=0$ wave-function components and so shifts radii by only $\sim 2.7\%$ even for the $L=1$ ($A=7$) nuclei. The $\Sigma'_1$ and $\Sigma''_1$ responses are thus effectively symmetry related even without isolating the $\mathcal{L}=0$ term.

\begin{figure}[t]
\centering
\includegraphics[width=\columnwidth]{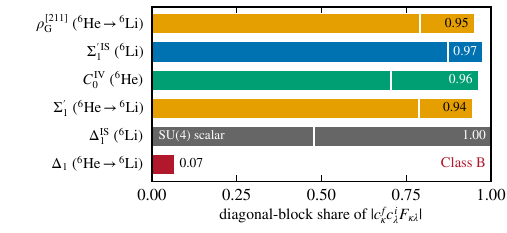}
\caption{Fraction of the block contributions $|c^f_\kappa c^i_\lambda F_{\kappa\lambda}|$ of Eq.~\eqref{eq:filter} of the $A=6$ responses carried by diagonal $\mathrm{SU}(4)$ blocks, integrated over $q\le400$~MeV/$c$ ($\mathrm{N}^2\mathrm{LO}_{\mathrm{sat}}$, $N_{\max}=8$, four lowest irreps). \textit{Class A} operators are colored by sector as in Fig.~\ref{fig:schematic} ($\mathrm{S}$ blue, $\mathrm{T}$ green, $\mathrm{G}$ orange), the isoscalar convection current (an $\mathrm{SU}(4)$ scalar) in gray, and the \textit{Class B} convection current $\Delta_1$ in dark red. The white tick marks the contribution of the $1\to1$ block alone. $\mathrm{NNLO}_{\mathrm{opt}}$ ($N_{\max}=10$) gives the same pattern.}
\label{fig:block_fraction}
\end{figure}
\begin{figure}[t]
\centering
\includegraphics[width=\columnwidth]{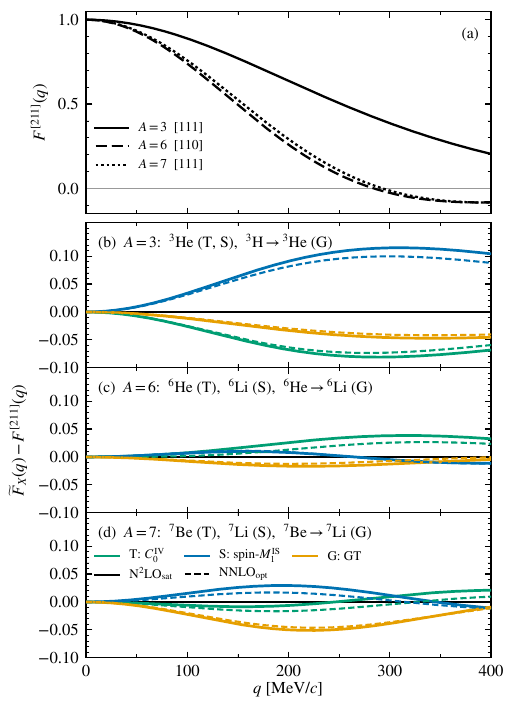}
\caption{(a) $\mathrm{SU}(4)$ adjoint form factor $F^{[211]}$ in the dominant irrep of $A=3$, $6$, and $7$ ($\mathrm{N}^2\mathrm{LO}_{\mathrm{sat}}$). Lines show the mean of the normalized $\mathcal{L}=0$ GT, spin-$M_1^{\mathrm{IS}}$, spin-$M_1^{\mathrm{IV}}$, and $C_0^{\mathrm{IV}}$ form factors. (b)-(d) Differences between the normalized form factors of the full wave functions and $F^{[211]}$ for the three sectors: $\mathrm{T}$ ($C_0^{\mathrm{IV}}$), $\mathrm{S}$ (spin-$M_1^{\mathrm{IS}}$), and $\mathrm{G}$ (GT). Solid: $\mathrm{N}^2\mathrm{LO}_{\mathrm{sat}}$; dashed: $\mathrm{NNLO}_{\mathrm{opt}}$, each relative to its own $F^{[211]}$.}
\label{fig:formfactors}
\end{figure}

\textit{Methods}---\emph{Nuclear states.} We use the no-core shell model (NCSM)~\cite{bar13} as implemented in JupiterNCSM~\cite{dja22,lim26_1} (version 2.0.0~\cite{JupiterNCSM_v2.0.0}). The many-body basis consists of harmonic-oscillator Slater-determinant states, truncated at $N_{\max}$ total oscillator excitations. Results are presented for the $\mathrm{N}^2\mathrm{LO}_{\mathrm{sat}}$~\cite{eks15} and $\mathrm{NNLO}_{\mathrm{opt}}$~\cite{eks13} Chiral Effective Field Theory ($\chi$EFT) interactions; the former includes three-nucleon forces. We use $\hbar\omega=20$~MeV, with truncations $N_{\max}=10$, $8$, and $6$ ($\mathrm{N}^2\mathrm{LO}_{\mathrm{sat}}$) and $12$, $10$, and $6$ ($\mathrm{NNLO}_{\mathrm{opt}}$) for $A=3$, $6$, and $7$, unless stated otherwise. (These model spaces do not reach sub-percent numerical convergence, but our conclusions are not affected.) Spurious center-of-mass motion is removed exactly from the operators following Ref.~\cite{nav21}.  

\emph{$\mathrm{SU}(4)$ decomposition.} Each eigenstate $|\Psi_{JM_T}\rangle$ is decomposed into $\mathrm{SU}(4)$ irreps by a Lanczos diagonalization of the quadratic Casimir of $\mathrm{SU}(4)$, $\mathcal{C}_2 = \bm{S}^2 + \bm{T}^2 + \sum_{\mu\nu}G^\dagger_{\mu\nu}G_{\mu\nu}$ with $|\Psi_{JM_T}\rangle$ as the pivot vector~\cite{joh15,mcc24,lim26_1}. This yields the eigenvalues of $\mathcal{C}_2$ and the amplitude of $|\Psi_{JM_T}\rangle$ in each eigenspace.

\emph{Form factors.} One-body transition densities $\psi^{f,i}_{J}(\alpha,\beta)$ obtained from the NCSM are folded with harmonic-oscillator reduced matrix elements~\cite{hax08} to form the many-body reduced matrix element,
\begin{equation}
  F(q) = \sum_{\alpha\beta}\psi^{f,i}_{J}(\alpha,\beta)\,
         \langle\alpha\|\mathcal{O}_J(q)\|\beta\rangle.
  \label{eq:fold}
\end{equation}
Writing each eigenstate as the sum of its normalized components in the eigenspaces of $\mathcal{C}_2$, $|\Psi\rangle=\sum_\lambda c_\lambda|\Phi_\lambda\rangle$, defines amplitudes $c^i_\lambda=\langle\Phi^i_\lambda|\Psi_i\rangle$ and $c^f_\kappa=\langle\Phi^f_\kappa|\Psi_f\rangle$, with $i$ and $f$ denoting the initial and final nuclear states of the transition\footnote{$|\Phi_\lambda\rangle$ is in general not a single basis state but a superposition of many $A$-nucleon states, each of which has different $S$, $L$, $T$, and orbital wave functions, but transforms as a particular $\mathrm{SU}(4)$ irrep, see Ref.~\cite{lim26_1} for details.}, in terms of which
\begin{align}
  F(q) &= \sum_{\kappa\lambda}c^f_\kappa\,c^i_\lambda\,F_{\kappa\lambda}(q), \nonumber \\ F_{\kappa\lambda}(q) &\equiv \langle\Phi^f_\kappa\|\mathcal{O}_J(q)\|\Phi^i_\lambda\rangle.
  \label{eq:filter}
\end{align}
Each component has the angular momentum of its eigenstate ($\mathcal{C}_2$ commutes with $\bm{J}$), so $F_{\kappa\lambda}$ follows from Eq.~\eqref{eq:fold} with transition densities for $\mathcal{C}_2$ eigenstates. The bare $F_{\kappa\lambda}$ show the block structure of the operator, independent of the irrep population, and the weighted terms $c^f_\kappa c^i_\lambda F_{\kappa\lambda}$ give each $\mathcal{C}_2$-block's contribution to the response. For the \textit{Class A} operators in $A=6$, the four lowest irreps reproduce the full form factors within $1\%$.

\emph{Radii.} Mean-square radii follow from $\widetilde{F}(q) = 1 - \tfrac{1}{6}\langle r^2\rangle q^2 + \mathcal{O}(q^4)$; we fit $1-\widetilde{F}$ with $aq^2+bq^4$ for $q\le40$~MeV/$c$, and the resulting radii agree with finite differences at $q=0$ to within $2\times10^{-4}$~fm, negligible compared with $\mathrm{SU}(4)$-breaking effects.


\textit{Results}---Figure~\ref{fig:block_fraction} demonstrates that the matrix elements of \textit{Class A} operators are dominated by supermultiplet-diagonal transitions, while those of non-scalar \textit{Class B} operators are not. For each response we compute the fraction of its $q$-integrated modulus coming from the diagonal terms $|c^f_\kappa c^i_\kappa F_{\kappa\kappa}|$ in Eq.~(\ref{eq:filter}). 
In the adjoint monopole density these diagonal terms carry 95\% of this sum. 
The operators $\Sigma'_1$ (which enters the transverse axial-electric and vector-magnetic responses) and $C_0^{\mathrm{IV}}$ show the same structure. In contrast, the charge-changing convection current ($\Delta_1$) connects irreps strongly: its response is built almost entirely from off-diagonal terms. Meanwhile, the isoscalar convection current of $^6$Li (an $\mathrm{SU}(4)$ scalar) is block-diagonal.

\begin{figure*}[t]
\centering
\includegraphics[width=\textwidth]{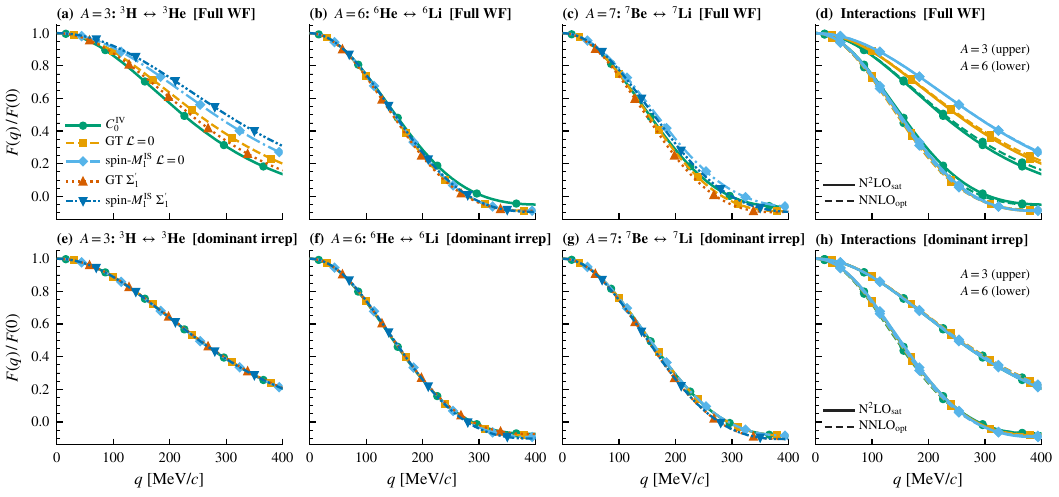}
\caption{Normalized electroweak form factors $\widetilde{F}(q) = F(q)/F(0)$ for the full wave functions (Full WF, top row) and the dominant irrep (bottom row). Panels (a, e), (b, f), and (c, g): $A=3$ ($^3\mathrm{H} \leftrightarrow {}^3\mathrm{He}$), $A=6$ ($^6\mathrm{He} \leftrightarrow {}^6\mathrm{Li}$), and $A=7$ ($^7\mathrm{Be} \leftrightarrow {}^7\mathrm{Li}$) with $\mathrm{N}^2\mathrm{LO}_{\mathrm{sat}}$. Channels: $C_0^{\mathrm{IV}}$, the $\mathcal{L}=0$ GT and spin-$M_1^{\mathrm{IS}}$ operators, and the bare $\Sigma'_1$ GT and spin-$M_1^{\mathrm{IS}}$ multipoles. Panels (d, h) compare $\mathrm{N}^2\mathrm{LO}_{\mathrm{sat}}$ (solid) and $\mathrm{NNLO}_{\mathrm{opt}}$ (dashed) for the $\mathcal{L}=0$ channels in $A=3$ (upper; $N_{\max}=10$ and $12$) and $A=6$ (lower; $N_{\max}=8$ and $10$).}
\label{fig:su4_ew_equivalence}
\end{figure*}

Within a supermultiplet, the \textit{Class A} operators remain algebraically connected at finite momentum transfer. This leads to the shape equivalence of Eq.~\eqref{eq:shapelock}, which we verify numerically in Fig.~\ref{fig:formfactors}. In the dominant irrep, the $\mathcal{L}=0$ $\mathrm{G}$ and $\mathrm{S}$ form factors collapse onto the $\mathrm{T}$ form factor for $A=3$, $6$, and $7$, i.e., onto a single $F^{[211]}$ [panel (a)], as required by Eq.~\eqref{eq:shapelock}. Their spread stays below $0.02$ up to $400$~MeV/$c$ (End Matter, Fig.~\ref{fig:su4_ew_equivalence}, bottom row). 

Panels (b)-(d) show the deviations of the physical one-body form factors from $F^{[211]}$. Here we see the combined effect of admixed irreps, the $\mathcal{L}=2$ term of $\Sigma'_1$, and, in $A=6$, the small orbital differences between $^6$He and $^6$Li (End Matter, Fig.~\ref{fig:su4_ew_equivalence}, top row). In $^6$Li, the dominant-irrep component also has an $L=2$ part, whereas $^6$He in its dominant irrep is pure $L=0$.

\begin{figure}[t]
\centering
\includegraphics[width=\columnwidth]{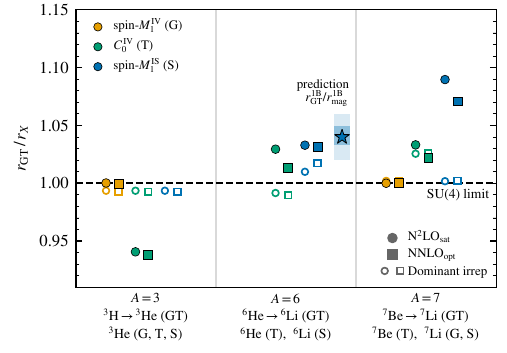}
\caption{Ratio of the Gamow--Teller radius to the radii of the other adjoint channels for $A=3$, $6$, and $7$ (bare $\Sigma'_1$ multipoles). Filled symbols: full wave functions. Open symbols: dominant irrep. Circles: $\mathrm{N}^2\mathrm{LO}_{\mathrm{sat}}$; squares: $\mathrm{NNLO}_{\mathrm{opt}}$. The dashed line is the $\mathrm{SU}(4)$ limit of the $\mathcal{L}=0$ operators. The star is the prediction $r^{\rm 1B}_{\mathrm{GT}}/r^{\rm 1B}_{\rm mag} = 1.04 \pm 0.02$ for $^6$He/$^6$Li (light band; dark band: range of all calculations). The $A=3$ spin-$M_1^{\mathrm{IS}}$ ratios ($1.28$ and $1.26$) lie outside the range shown.}
\label{fig:ratios}
\end{figure}

Figure~\ref{fig:ratios} focuses on form-factor ratios near $q=0$, displaying radii ratios $r_{\mathrm{GT}}/r_X$, which equal unity in the $\mathrm{SU}(4)$ limit. The radii themselves are listed in Table~\ref{tab:su4_triality_master_radii} of the End Matter. In the dominant irrep (open symbols), all ratios lie within $2.6\%$ of unity. The largest departure, for $C_0^{\mathrm{IV}}$ in $A=7$, comes from the $\mathcal{L}=2$ term of $\Sigma'_1$: it vanishes between the $L=0$ states that dominate $A=3$ and $6$, but in the $L=1$ states of $A=7$ it raises the Gamow--Teller and spin-$M_1^{\mathrm{IS/IV}}$ radii together by $2.7\%$ (that particular ratio stays at one). Considering only $\mathcal{L}=0$ operators produces dominant-irrep radii that are all within $1.3\%$ of $r_{\mathrm{GT}}$.

In the full wave functions (filled symbols), the two $\mathrm{G}$ components, GT and spin-$M_1^{\mathrm{IV}}$, have equal radii to better than $0.1\%$ in $A=3$ and $7$. This is a consequence of isospin, not $\mathrm{SU}(4)$; $\mathrm{SU}(4)$ connects different sectors. The GT and $C_0^{\mathrm{IV}}$ ($\mathrm{T}$) radii agree to within about $3\%$ in $A=6$ and $7$ and $6$-$7\%$ in $A=3$ (not converged in $N_{\max}$). The larger $A=3$ value is driven by $C_0^{\mathrm{IV}}$; in the dominant irrep the proton and neutron distributions of $^3$He coincide, whereas in the full wave function the protons are $0.15$~fm more extended. The isoscalar radius barely changes, but $(r^{\mathrm{IV}}_{C_0})^2=2r_p^2-r_n^2$ magnifies the difference such that $r^{\mathrm{IV}}_{C_0}$ grows twice as much as $r_{\mathrm{GT}}$. In $A=6$, where the transition connects different isospin multiplets, the spin-$M_1^{\mathrm{IS}}$ ($\mathrm{S}$) radius agrees as well. Only the spin-$M_1^{\mathrm{IS}}$ radii of the mirror systems deviate more (ratios $1.07$-$1.09$ in $A=7$ and $1.26$-$1.28$ in $A=3$). Here symmetry breaking shifts the two radii in opposite directions; relative to the dominant irrep, $r_{\mathrm{GT}}$ grows by $5$-$6\%$ ($2\%$), while the spin-$M_1^{\mathrm{IS}}$ radius shrinks by $17$-$18\%$ ($5$-$6\%$) in $A=3$ ($A=7$).

In the $\mathrm{SU}(4)$ limit, the $^6$He$\to{}^6$Li Gamow--Teller radius equals the spin-$M_1^{\mathrm{IS}}$ radius of the $^6$Li ground state ($J=1$, $T=0$). In our calculations the two radii are strongly correlated. For the $\mathcal{L}=0$ operators, the full wave functions give a ratio of $1.011$-$1.014$ across both interactions and all model spaces, although the Gamow--Teller radius itself varies by $8\%$. If the form factors are not ${\cal L}$-decomposed the ratio rises to $1.030$-$1.035$ (End Matter), almost entirely through the $\mathcal{L}=2$ part of $\Sigma'_1$. The measured form factor also contains the convection current, which vanishes between $L=0$ components but supplies $7$-$9\%$ of the one-body magnetic form factor at $q=0$ through the $L\neq0$ components of $^6$Li. This raises the ratio by $1.7$-$2.7\%$. Meanwhile, the physical current weights the nucleon spins by their magnetic moments, lowering the ratio by $1.1$-$1.7\%$. Ultimately, the ratio of the one-body Gamow--Teller radius to the full one-body point magnetic radius $r^{\rm 1B}_{\rm mag}$ of $^6$Li is $1.033$-$1.049$ across all our calculations.

Searches for the Fierz interference term induced by tensor currents beyond the Standard Model are targeting a measurement of the $^6$He $\beta$ spectrum at per-mil precision~\cite{byr23,gli22_1}.
The Gamow--Teller radius determines terms of second order in the lepton momentum $q$ and so affects the standard-model background for this spectrum at the $10^{-3}$ level~\cite{kin23}.
Reference~\cite{kin23} pointed out that the $8$-$13\%$ uncertainty in $r_{\mathrm{GT}}$ produces an  uncertainty $\sim 10^{-4}$ in the ${}^6$He beta-decay spectrum. 
Assigning a conservative $50\%$ uncertainty to the $\mathrm{SU}(4)$ symmetry-breaking contributions discussed in this work, we obtain $r^{\rm 1B}_{\mathrm{GT}} = (1.04 \pm 0.02)\,r^{\rm 1B}_{\rm mag}$ (star in Fig.~\ref{fig:ratios}), with an interval that covers all our calculations and is $2.5$ times wider than their spread. Existing data on the ${}^6$Li magnetic radius~\cite{gol63,ran66,ber82} mostly predate 1984 and have large uncertainties~\cite{cha24}, but a high-precision measurement of the ${^6}$Li magnetic form factor at low $q$ would determine the one-body Gamow--Teller radius to about $2\%$ plus the experimental error, markedly reducing the uncertainty quoted in  Ref.~\cite{kin23}.
In $\chi$EFT axial two-body currents enter at order $Q^2 $ ($Q^3$) relative to leading in the EFT with (without) an explicit $\Delta(1232)$ degree of freedom~\cite{kre20,bar16,kin23}, with $Q \approx 0.3$ the EFT expansion parameter. Meanwhile, two-body currents enter the isoscalar magnetic form factor at relative $O(Q^3)$. The leading one-body matrix elements are not suppressed, since the transition is $\mathrm{SU}(4)$ allowed~\cite{lim26_1}, so two-body currents are small in both cases~\cite{kin23}. 
Adding them will therefore not markedly change our overall precision estimate, but would precisely constrain the radii entering the evaluation of the recoil-order nuclear-structure corrections to the $^6$He decay.

\textit{Conclusions}---Emergent Wigner $\mathrm{SU}(4)$ symmetry ties the vector and axial-vector responses of light nuclei to a single form factor. At any momentum transfer, the isovector charge, the Fermi operator, and the orbital-scalar parts of the spin-$M_1^{\mathrm{IS/IV}}$ and Gamow--Teller operators form one $\mathrm{SU}(4)$ adjoint monopole density. Its form factor $F^{[211]}$ describes how the $\mathrm{SU}(4)$ charges are distributed radially in the nucleus. For the dominant supermultiplet in each of the $A=3$, $6$, and $7$ systems the form factor computed using the NCSM follows $F^{[211]}$ up to $q\approx400$~MeV/$c$. Within this supermultiplet the radii of the orbital-scalar operators are all within $1.3\%$ of $r_{\mathrm{GT}}$. Among the one-body electroweak multipole operators~\cite{don79,hax08}, those of \textit{Class A} contain components of the adjoint monopole density and remain approximately block-diagonal in $\mathrm{SU}(4)$ irreps, whereas non-scalar \textit{Class B} operators drive transitions between irreps and are sensitive to symmetry breaking.

Our results contribute to several highly scrutinized research fields. For instance, in neutrino--nucleus scattering, CVC and unpolarized electron-scattering data constrain only the vector part of the weak response~\cite{alv18}. Here we have shown that because $F^{[211]}$ is shared by vector and axial responses, electromagnetic data also constrain one-body axial responses, at least for the symmetry-protected responses of light nuclei.
Another application is the $^6$He Gamow--Teller radius, which we predict to be $(4\pm2)\%$ larger than the one-body magnetic radius of $^6$Li.
High-precision measurements of the low-$q$ magnetic form factor of $^6$Li, combined with the $M1$ width of its $3.56$~MeV state, would then constrain both the leading $\mathcal{O}(q)$ and the leading $\mathcal{O}(q^2)$ corrections to the $^6$He $\beta$ spectrum used in searches for physics beyond the Standard Model~\cite{gli22_1}.
Finally, the distinction between \textit{Class A} and \textit{Class B} operators---and the separation of $\mathrm{SU}(4)$-protected responses from those driven by symmetry-breaking admixtures---has not yet been built into Bayesian models of EFT truncation errors~\cite{fur15,mel17}. Developing error models for observables shaped by emergent symmetries is an important topic for future work.
\vspace{0.1cm}
\begin{acknowledgments}
We acknowledge several helpful discussions with Tor R. Djärv regarding the numerical part of this project. This research was supported by the Swedish Research Council via a Tage Erlander Visiting Professorship, Grant No.\ 2022-00215 (D.R.P.\ and SS.L.M.), and a Research Project Grant, No.\ 2021-04507 (C.F.\ and SS.L.M.) as well as by the US Department of Energy under contract No.\ DE-FG02-93ER40756 (D.R.P.). This project is supported in part by the National Science Foundation FRHTP program under award No.  PHY-2402275 (D.R.P.). SS.L.M. acknowledges travel support from the Barbro Osher Endowment, Scholarship No.\ 90401181, and the Royal Swedish Academy of Sciences, Scholarship No.\ PH2024-0090. SS.L.M. is grateful for hospitality provided by Oak Ridge National Laboratory (ORNL) and Ohio University, where part of this work was done. The computations and data handling were partly enabled by resources provided by the National Academic Infrastructure for Supercomputing in Sweden (NAISS), partially funded by the Swedish Research Council through grant agreement No.\ 2022-06725. A portion of the computation for this work was performed on the University of Tennessee Infrastructure for Scientific Applications and Advanced Computing (ISAAC) computational resources. Gemini 3.1 Pro (Google) and Claude Opus 5.5 (Anthropic) assisted in developing the results of this project. The general pipeline worked as follows: Derivations typically started with an idea and a sketch of the proof with details constructed through iterative discussion with the two AIs. All outputs were reviewed for accuracy and queried where necessary until the derivation was complete. The AI tools also assisted in developing and debugging the code that computes the one-body matrix elements, as well as the analysis and plotting scripts. We also used these tools to aid figure design and check the generalized Wigner--Eckart theorem. We directed all AI-assisted work and validated it against independent results (Table~\ref{tab:su4_limits}, Ref.~\cite{lim26_1}).

\end{acknowledgments}

\bibliography{mybibfile}

\clearpage
\section{End Matter}

\paragraph{Definition of the seven basic one-body multipole operators.}
In the conventions of Refs.~\cite{don79,hax08}, the single-nucleon space-spin multipole operators obtained by expanding the electroweak charge and current operators through order $|\bm{p}|/M_N$ are constructed from spherical Bessel functions $j_L(qx)$, spherical harmonics $Y_{JM}(\Omega_x)$, and vector spherical harmonics $\bm{Y}^M_{JL1}(\Omega_x) = [Y_L(\Omega_x)\otimes\bm{e}_1]^M_J$ via $\bm{M}^M_{JL}(q\bm{x}) \equiv j_L(qx)\,\bm{Y}^M_{JL1}(\Omega_x)$ (their charge multipole $M_J$ is denoted $C_J$ here):
\begin{align}
  C^M_J(q\bm{x}) &\equiv j_J(qx)\,Y_{JM}(\Omega_x), \nonumber \\
  \Delta^M_J(q\bm{x}) &\equiv \bm{M}^M_{JJ}(q\bm{x})\cdot\frac{1}{q}\bm{\nabla}, \nonumber \\
  \Delta'^M_J(q\bm{x}) &\equiv -i\left[\frac{1}{q}\bm{\nabla}\times\bm{M}^M_{JJ}(q\bm{x})\right]\cdot\frac{1}{q}\bm{\nabla} \nonumber \\
  &= [J]^{-1}\!\big[-\sqrt{J}\,\bm{M}^M_{J,J+1} + \sqrt{J{+}1}\,\bm{M}^M_{J,J-1}\big]\!\cdot\frac{1}{q}\bm{\nabla}, \nonumber \\
  \Sigma^M_J(q\bm{x}) &\equiv \bm{M}^M_{JJ}(q\bm{x})\cdot\bm{\sigma}, \nonumber \\
  \Sigma'^M_J(q\bm{x}) &\equiv -i\left[\frac{1}{q}\bm{\nabla}\times\bm{M}^M_{JJ}(q\bm{x})\right]\cdot\bm{\sigma} \nonumber \\
  &= [J]^{-1}\!\big[-\sqrt{J}\,\bm{M}^M_{J,J+1} + \sqrt{J{+}1}\,\bm{M}^M_{J,J-1}\big]\!\cdot\bm{\sigma}, \nonumber \\
  \Sigma''^M_J(q\bm{x}) &\equiv \left[\frac{1}{q}\bm{\nabla} C^M_J(q\bm{x})\right]\cdot\bm{\sigma} \nonumber \\
  &= [J]^{-1}\!\big[\sqrt{J{+}1}\,\bm{M}^M_{J,J+1} + \sqrt{J}\,\bm{M}^M_{J,J-1}\big]\!\cdot\bm{\sigma}, \nonumber \\
  \Omega^M_J(q\bm{x}) &\equiv C^M_J(q\bm{x})\,\bm{\sigma}\cdot\frac{1}{q}\bm{\nabla}.
  \label{eq:seven_ops}
\end{align}
For $J=1$, $\sqrt{\tfrac{2}{3}}\,\Sigma'_{1\mu}+\tfrac{1}{\sqrt{3}}\,\Sigma''_{1\mu}=\bm{M}^{\mu}_{10}\cdot\bm{\sigma}=j_0(qx)\,Y_{00}(\Omega_x)\,\sigma_\mu$ is the $\mathcal{L}=0$ term of the main text, proportional to $\rho^{[211]}_{(1\mu)(00)}$ [$\rho^{[211]}_{(1\mu)(1\nu)}$] in the isoscalar (isovector) channel; the orthogonal combination $\sqrt{\tfrac{2}{3}}\,\Sigma''_{1\mu}-\tfrac{1}{\sqrt{3}}\,\Sigma'_{1\mu}=\bm{M}^{\mu}_{12}\cdot\bm{\sigma}$ is the $\mathcal{L}=2$ term.

\paragraph{General form of the GWE theorem.}
For a one-body operator obtained by coupling the spin index $\mu$ of $\Gamma^{[211]}_{(S_{\rm op}\mu)(T_{\rm op}\nu)}$ with an orbital tensor of rank $\mathcal{L}$, $S_{\rm op}\otimes\mathcal{L}\to J_{\rm op}$, the reduced matrix elements between states of supermultiplets $[f_i]$ and $[f_f]$ with $S\otimes L\to J$ coupling are
\begin{align}
&\big\langle [f_f]S_fT_fM_{T_f}\,L_f\,J_f\big\|\mathcal{O}^{[211]}_{(S_{\rm op}\mathcal{L})J_{\rm op}\,(T_{\rm op}\nu)}\big\|[f_i]S_iT_iM_{T_i}\,L_i\,J_i\big\rangle \nonumber \\
&\quad= [J_f]\,\mathrm{CG}_T\,[S_f][L_f][J_i][J_{\rm op}]
\begin{Bmatrix} S_i & L_i & J_i\\ S_{\rm op} & \mathcal{L} & J_{\rm op}\\ S_f & L_f & J_f\end{Bmatrix}
W\,\mathcal{R}_{\mathcal{L}}(q),
\label{eq:gwe_general}
\end{align}
with $\mathrm{CG}_T$ and $W$ defined below Eq.~(\ref{eq:gwe}) of the main text. The reduced matrix element $\mathcal{R}_{\mathcal{L}}(q)$ depends only on $q$, $[f_i]$, $[f_f]$, and the orbital wave functions, not on $S$, $T$, $J$, or $M_T$. For $\mathcal{L}=0$, $J_{\rm op}=S_{\rm op}$ and Eq.~(\ref{eq:gwe_general}) is Eq.~(\ref{eq:gwe}), with $\mathcal{R}_0=\mathcal{R}$.

\paragraph{Reduced $\mathrm{SU}(4)$ Wigner coefficients.}
The constants $c_X\equiv[J_f]\,\mathrm{CG}_T\,\mathcal{G}\,W$ of the main text combine the isospin Clebsch--Gordan coefficient, the $9j$ recoupling factor $\mathcal{G}$, and the reduced Wigner coefficient $W$; all three depend on the sector through $(S_{\rm op},T_{\rm op})$ and the states it connects. We denote by $W_X$ the Wigner coefficient of sector $X$ and obtain it from the algebraic expressions of Hecht and Pang~\cite{hec69}. The coefficients refer to the standard tensor $T^{[211]}=\sum_k\Gamma^{[211]}(k)$ of their Table~I, whose components are the generators $T^{[211]}_{(1\mu)(00)}=S_\mu$, $T^{[211]}_{(00)(1\nu)}=T_\nu$, and $T^{[211]}_{(1\mu)(1\nu)}=-G_{\mu\nu}$. The sign of the $\mathrm{G}$ sector follows from their phase convention for the $\mathrm{SU}(4)$ step operators and affects only the sign of the $\mathrm{G}$-sector coefficients. For the $[110]$ supermultiplet of $A=6$, $W_{\mathrm{G}}=\sqrt{3/5}$ for $(S,T)=(0,1)\to(1,0)$, and $W_{\mathrm{S}}=W_{\mathrm{T}}=\sqrt{2/5}$ within the $(1,0)$ and $(0,1)$ states, respectively. For the $[111]$ supermultiplet of $A=3$ and $A=7$, $W_{\mathrm{G}}=\sqrt{3/5}$ and $W_{\mathrm{S}}=W_{\mathrm{T}}=\sqrt{1/5}$. Together with the GWE theorem, these coefficients reproduce exactly the analytic GT reduced matrix elements of Ref.~\cite{lim26_1}.

\begin{table}[b]
\centering
\caption{Exact $\mathrm{SU}(4)$-limit values at $q=0$ of the $\mathcal{L}=0$ adjoint form factors of the main text, and the resulting $|F_{\mathrm{GT}}/F^{\mathrm{IS}}_{M_1}|$ ratios. Both are Edmonds reduced matrix elements.}
\label{tab:su4_limits}
\begin{tabular}{lccc}
\hline\hline
System & $|F^{\mathrm{SU(4)}}_{\mathrm{GT}}|$ & $|F^{\mathrm{IS,SU(4)}}_{M_1}|$ & $|F_{\mathrm{GT}}/F^{\mathrm{IS}}_{M_1}|$ \\
\hline
$A=3$ & $\sqrt{6}/2\sqrt{\pi} = 0.691$  & $\sqrt{6}/4\sqrt{\pi} = 0.345$ & $2$ \\
$A=6$ & $\sqrt{6}/2\sqrt{\pi} = 0.691$  & $\sqrt{6}/2\sqrt{\pi} = 0.691$ & $1$ \\
$A=7$ & $\sqrt{15}/3\sqrt{\pi} = 0.728$ & $\sqrt{15}/6\sqrt{\pi} = 0.364$ & $2$ \\
\hline\hline
\end{tabular}
\end{table}

\begin{table*}[ht]
\centering
\caption{Root-mean-square point radii (in fm) of the multipoles $\Sigma'_1$ and $C_0$ in the three $\mathrm{SU}(4)$ adjoint sectors: GT ($r_{\mathrm{GT}}$) and spin-$M_1^{\mathrm{IV}}$ ($r^{\mathrm{IV}}_{M_1}$) for $\mathrm{G}$, spin-$M_1^{\mathrm{IS}}$ ($r^{\mathrm{IS}}_{M_1}$) for $\mathrm{S}$, and $C_0^{\mathrm{IV}}$ ($r^{\mathrm{IV}}_{C_0}$) for $\mathrm{T}$, evaluated in $^3\mathrm{He}$, $^6\mathrm{He}$, $^7\mathrm{Be}$ (charge) and $^3\mathrm{He}$, $^6\mathrm{Li}$, $^7\mathrm{Li}$ (spin). $N_{\max}$ is given for $\mathrm{N}^2\mathrm{LO}_{\mathrm{sat}}$ and $\mathrm{NNLO}_{\mathrm{opt}}$.}
\label{tab:su4_triality_master_radii}
\begin{tabular}{llcccc@{\hspace{2em}}cccc}
\hline\hline
 & & \multicolumn{4}{c}{$\mathrm{N}^2\mathrm{LO}_{\mathrm{sat}}$} & \multicolumn{4}{c}{$\mathrm{NNLO}_{\mathrm{opt}}$}\\
System ($N_{\max}$) & State & $r_{\mathrm{GT}}$ & $r^{\mathrm{IV}}_{M_1}$ & $r^{\mathrm{IS}}_{M_1}$ & $r^{\mathrm{IV}}_{C_0}$ & $r_{\mathrm{GT}}$ & $r^{\mathrm{IV}}_{M_1}$ & $r^{\mathrm{IS}}_{M_1}$ & $r^{\mathrm{IV}}_{C_0}$\\
\hline
$A=3$ (10, 12) & Dominant irrep & 1.6658 & 1.6766 & 1.6766 & 1.6766 & 1.6557 & 1.6680 & 1.6680 & 1.6680 \\
               & Full WF        & 1.7615 & 1.7611 & 1.3777 & 1.8726 & 1.7454 & 1.7467 & 1.3884 & 1.8601 \\[4pt]
$A=6$ (8, 10)  & Dominant irrep & 2.6231 & --     & 2.5973 & 2.6454 & 2.7134 & --     & 2.6667 & 2.7416 \\
               & Full WF        & 2.6642 & --     & 2.5793 & 2.5879 & 2.7431 & --     & 2.6585 & 2.7070 \\[4pt]
$A=7$ (6, 6)   & Dominant irrep & 2.5776 & 2.5723 & 2.5730 & 2.5133 & 2.5433 & 2.5377 & 2.5380 & 2.4791 \\
               & Full WF        & 2.6305 & 2.6305 & 2.4140 & 2.5460 & 2.5834 & 2.5826 & 2.4123 & 2.5281 \\
\hline\hline
\end{tabular}
\end{table*}

\paragraph{Absolute normalization and the $q=0$ ratios.}
All form factors are reduced matrix elements in the convention of Edmonds~\cite{edm57}, as are the single-particle matrix elements~\cite{don79,hax08}. The NCSM one-body transition density of Eq.~(\ref{eq:fold}), $\psi^{f,i}_J(\alpha,\beta) = -\langle J_f\|(a^\dagger_\alpha\tilde{a}_\beta)_J\|J_i\rangle/\sqrt{2J{+}1}$, is reduced in $J$ in the proton--neutron basis~\cite{nav21}. These proton--neutron form factors are then converted to the isospin formalism by taking appropriate linear combinations for the isoscalar and isovector cases.
Combining the GWE theorem with $\mathcal{R}(0)=\sqrt{\mathcal{C}_2/4\pi}$, which follows from $\langle[f]\|T^{[211]}\|[f]\rangle=\sqrt{\mathcal{C}_2}$~\cite{hec69} for identical orbital states, yields the closed-form $q=0$ predictions of Table~\ref{tab:su4_limits}. With $t_\pm=(\tau_x\pm i\tau_y)/2$, the Gamow--Teller operator is $\sigma_\mu t_\pm=\pm\sqrt{2}\,\Gamma^{[211]}_{(1\mu)(1,\pm1)}$, so $F_{\mathrm{GT}}(0)=\pm\sqrt{2}\,c_{\mathrm{G}}\mathcal{R}(0)$, whereas $F^{\mathrm{IS}}_{M_1}(0)=c_{\mathrm{S}}\mathcal{R}(0)$. 
In the dominant irrep, the computed ratios reproduce the last column of Table~\ref{tab:su4_limits} to $0.01\%$ in $A=3$ and $A=7$;
the residual measures isospin breaking between the mirror wave functions. In $A=6$ the computed ratio exceeds the SU(4) prediction by $5.4\%$. This is consistent with an $L=2$ part of about $7\%$ in the $^6$Li component, which lowers $F^{\mathrm{IS}}_{M_1}(0)$ more than $F_{\mathrm{GT}}(0)$.

\paragraph{Radii.}
Table~\ref{tab:su4_triality_master_radii} lists the root-mean-square point radii $r_X\equiv\langle r_X^2\rangle^{1/2}$ for both interactions. In the $\mathrm{SU}(4)$ limit they coincide for the $\mathcal{L}=0$ operators; the non-${\cal L}$-decomposed $\Sigma'_1$ radii also contain the $\mathcal{L}=2$ term.

\paragraph{Model-space convergence and radii correlations.}
Absolute radii are not converged at the reference truncations ($\hbar\omega=20$~MeV). For $A=6$, the GT radius grows by about $0.07$~fm per step in $N_{\max}$ for both interactions, from $2.53$ to $2.66$~fm ($\mathrm{N}^2\mathrm{LO}_{\mathrm{sat}}$, $N_{\max}=4$-$8$) and from $2.60$ to $2.74$~fm ($\mathrm{NNLO}_{\mathrm{opt}}$, $N_{\max}=6$-$10$); for $A=3$, radii change by up to $0.05$~fm between $N_{\max}=8$ and $10$. The ratios between channels, which are tests of $\mathrm{SU}(4)$, are much more stable. In $A=6$, $r_{\mathrm{GT}}/r^{\mathrm{IS}}_{M_1} = 1.035$, $1.031$, $1.033$ at $N_{\max}=4$, $6$, $8$ ($\mathrm{N}^2\mathrm{LO}_{\mathrm{sat}}$) and $1.030$, $1.033$, $1.032$ at $N_{\max}=6$, $8$, $10$ ($\mathrm{NNLO}_{\mathrm{opt}}$). Across both interactions and all model spaces, the GT radius varies by $8\%$ while this ratio stays within $1.030$-$1.035$. 
In $A=3$, $r_{\mathrm{GT}}/r^{\mathrm{IV}}_{M_1}$ lies within $0.3\%$ of unity for all $N_{\max}\ge4$ and both interactions. For $C_0^{\mathrm{IV}}$, $r_{\mathrm{GT}}/r^{\mathrm{IV}}_{C_0}$ is $1.031$, $1.031$, $1.029$ ($\mathrm{N}^2\mathrm{LO}_{\mathrm{sat}}$) and $1.018$, $1.016$, $1.013$ ($\mathrm{NNLO}_{\mathrm{opt}}$) in $A=6$ at the same truncations, and $0.955$, $0.948$, $0.941$ in $A=3$ at $N_{\max}=6$, $8$, $10$ ($\mathrm{N}^2\mathrm{LO}_{\mathrm{sat}}$) and $0.951$, $0.945$, $0.938$ at $N_{\max}=8$, $10$, $12$ ($\mathrm{NNLO}_{\mathrm{opt}}$). For $A=7$, calculations are limited to $N_{\max}=6$.

\end{document}